\documentclass[aps,prd,superscriptaddress,eqsecnum,amsfonts,showpacs,epsfig]{revtex4}
\usepackage{graphicx}
\usepackage{xcolor}
\usepackage{ulem}
\usepackage{subcaption}

\newcommand{\be}{\begin{equation}}
\newcommand{\ee}{\end{equation}}
\newcommand{\bea}{\begin{eqnarray}}
\newcommand{\eea}{\end{eqnarray}}

\newcommand{\nn}{\nonumber}
\newcommand{\ep}{i\epsilon}
\newcommand{\om}{\omega}

\begin{document}

\title{Dispersive analysis of the $\pi^+ \pi^-$  production at the CMD-3 experiment and
 the compatibility with  muon pair production  measurement by KLOE-2 and the pion form factor  by JLAB}
\author{Dimitrios Petrellis and Vladimir \v{S}auli}
\affiliation{Department of Theoretical Physics, Institute of Nuclear Physics Rez near Prague, CAS, Czech Republic  }

\begin{abstract}

The spectral function of the charged pion form factor was extracted from two data sets. The difference between the two sets is based on the presence or absence of the recent measurement of the $\pi^+ \pi^-$  production by CMD-3 . Although the CMD-3 data are largely incompatible with other recent measurements, no excess was found when the data were used for analytical continuation to the spacelike region and compared to JLaB-$\pi$ experiment.  Instead, both data sets provide a spacelike form factor that differs from the  expected behavior known from perturbative analyses of quantum chromodynamics (QCD).  A precise fit of the   spectral functions is provided, either based on he bare form factor and the full $e^+e^-\rightarrow \pi^+\pi^-$ as well . Furthermore, the extracted  pionic spectral functions  were used to obtain  the  QED running  charge. This was then   compared with  the KLOE-2 measurements. 

\end{abstract}


\maketitle


%
\section{Introduction}

The precise determination of  Hadronic Vacuum Polarization (HVP) represents an important  issue for comparison  between theory and experiment.
It is the source of  the largest systematic uncertainties in the determination of muonic magnetic moment  \cite{amu1,amu2,Abi2021}.  
To this end, the optical theorem and a few other non-perturbative methods  \cite{HGK2017,GFW2011}
 can be used to calculate the muon anomalous magnetic moment, $a_{\mu}$. 

A plethora of precise experiments conducted over the last two decades at BaBaR, Belle, BESS, and by the SND/CMD collaborations provided data on exclusive processes    $e^+e^-\rightarrow hadrons $ that are needed to determine  the HVP. However,   the associated data for the most important two-particle channel,$e^+e^-\to\pi^+\pi^-$,  exhibit significant mutual incompatibility when different experiments are compared. Based on the new generation experimental setup  CMD-3, the experimental group in Novosibirsk provided in \cite{CMD3} the measured data on the charged pion pair production cross section that deviates most significantly in the $\rho $ resonance region. This  suggests   an unknown  systematic error presented in measurement. While the difference in measured cross sections below $1  {\mbox GeV} $ was already significant, the difference between averaged world data and the data collected in  \cite{CMD3}  now exceeds ten standard deviations at the rho meson peak. It is worth mentioning that the statistical errors are essentially identical for all the aforementioned experiments. 
While the mutual incompatibility of the pion electroproduction cross section is striking, one might ask if it is possible to evaluate the quality of the data using knowledge from other related experiments. 
The main motivation behind this work is to examine how the aforementioned incompatibility affects the theoretical calculation of two related observables. First, using an assumed dispersion relation to check the compatibility of the measured pion form factors at spacelike kinematics. 
 To this end, we extract the spectral function of the pion charge form factor, perform an analytical continuation and compare to the JLAB-$\pi$ measurements \cite{JLAB2008,JLABb2008}. 
Second, we extract the HVP with and without CMD-3 data, comparing the theory to the KLOE-2 measurement\cite{KLOEdva} of the fine structure coupling constant. 
A third objective is to correct the numerical errors identified in the author's earlier calculation of the pion form factor \cite{sauliarxiv} .    

The paper is organized as follows:
 The next section provides the necessary theory for calculating the running charge. 
Section \ref{infl} explains the combination of data and error propagation.
 A separate section, section \ref{DRFOR} , discusses the preparation of the pion cross section for the HVP calculation and describes the extraction of the spectral function.  In this section, we also compare it to the JLAB-$\pi$ experiment.
 Section \ref{alfy} compares the calculated running QED coupling with the experiment and draws conclusions.

\section{Vacuum polarization calculation}

The fine structure constant is a different name for the  running QED charge $\alpha(s)$, it reads
\be \label{alfa}
\alpha(s)=\frac{\alpha}{1-\Pi(s)} \, ,
\ee
with $\alpha=\alpha(0)=1/137.0359991390$ and  the polarization function
$\Pi(s)$. Unlike the other two Standard Model charges, it is entirely included in the single correlator, the photon propagator. 

Experimentally, the square of the  running charge $\alpha(s)$ has been extracted  from the muon pairs production integral cross-section $\sigma_{\mu\mu}=\sigma(e^{+}e^{-}\to \mu^{+}\mu^{-}) $ with KLOE experimental cut  on polar scattering angle between $\mu^{-}$ and $e^{-}$ particles.

 The interesting structure observed in the function $\alpha(s)$ is aconsequence of interference  between two complex functions: the  HVP $\Pi_h$ and the leptonic vacuum polarization $\Pi_l$, composing the the polarization function $\Pi(s)=\Pi_l(s)+\Pi_h(s) $ in the Eq. \ref{alfa}.  The constant absorptive part of the leptonic polarization provides a background that constitutes a unique interference pattern that emerges in the vicinity of the meson resonance.

Leptonic contributions are  well known from perturbation theory \cite{ARBU1997}.  For the sake of completeness, we present the leptonic contribution to the vacuum polarization function below:
\be
\Pi_l(s)=\frac{\alpha}{\pi}\Pi_1(s)+\left(\frac{\alpha}{\pi}\right)^2\Pi_{2e}(s) \, \, ,
\ee
where one the loop contribution is
\bea \label{lepton}
\Pi_1(s)&=&\Pi_e(s)+\Pi_{\mu}(s)+\Pi_{\tau}(s)\, \, ;
\nn \\
\Pi_f(s)&=&-5/9-x_f/3+f(x_f)\,\, ; f=e,\mu,\tau \, ; 
\nn \\
f(x_f)&=&\frac{\beta_f}{6}(2+x_f) 
\left(\ln{\frac{1+\beta_{f}}{1-\beta_f}} -i\pi\right)\Theta(1-x_f)
\nn \\
&+&\frac{\beta_f}{3}(2+x_f)  \arctan \left({\frac{1}{\beta_f}}\right)\Theta(x_f-1)  \, \, ,
\eea
where $\beta_f=\sqrt{|1-x_f|}$ and $x_f=4m^2_f/s$.
Also the leading second order logarithmic term
\be
\Pi_{2e}(s)=\frac{1}{4}(\ln\frac{s}{m_e^2}-i\pi)+\zeta(3)-5/24 \, ,
\ee
is taken into account. For heavy quarks and large $q^2$ we employ perturbation theory with the extra factor $\alpha\rightarrow \alpha N_c e_q^2$ in the appropriate one-loop expression. 

The HVP function  $\Pi_h$ is  extracted from the total hadronic production $\sigma_h=\sigma_{tot}(e^+e^-\rightarrow hadrons) $
by using the following dispersion relation \cite{CABGAT1961,EIDJEg1995}:
\be \label{muf}
\Pi_h(s)=\frac{s}{4\pi^2\alpha}\int_{m_{\pi}^2}^{\infty}d\omega \frac{\sigma_h(\omega) \left[\frac{\alpha}{\alpha(\omega)}\right]^2}
{\omega-s+i\epsilon}\, .
\ee

The knowledge of this quantity is dependent upon a multitude of experimental measurements 
 of the hadronic exclusive  processes. i.e.
\be
\sigma_h=\sum_{i}\sigma_{h_i} \, \, ,
\ee
with $i=\pi\pi, KK, \pi\pi\pi, \pi \gamma,....$  noting that the photons emitted from the final hadronic states count as well.

The expression (\ref{muf}) represents a nonlinear integral equation  with a singular kernel. To date, several groups \cite{polar2,polar3,DAVIER2011} have  steadily  collected necessary data on $\sigma_h$   and  continuously provided a fresh  look at the $\Pi_h$ line-shape. In our case, the experimental group (e.g. for experimental data BaBar, SND, BESS) provided the so-called bare cross section -i.e. with vacuum polarization  extracted out  by the experimental group itself, while CMD-3 data represent the full cross section. Hence combined  data were unified by adding or removing VHP and we determine the inflated error for the bare  as well as for the full cross section, as defined and described  below. It is done for purpose of presentation of the spectral functions. Note that in principle  one can combine $\sigma_B$ and $\sigma_F$ by labeling the data points and thus add or remove VHP iteratively.   
  
 What follows is a description of the determination of $\sigma_h$ in our case. 
To this point, the narrow resonances, such as heavy quarkonia, have been substituted by their Breit-Wigner functions and PDG parameters.
The inclusion of the remaining part of the total cross-section, denoted by $d\sigma_h$, is performed numerically and constitutes the core of the method.
Since the kernel in Eq. (\ref{muf}) is singular, a straight use of experimental data to integrate the Eq.  (\ref{muf})
 would lead to uncontrolled numerical noise and loss of accuracy.
 
 To  control the principal value integration, we first construct an analytic fit of $\sigma_h$ as  the first step. Using this fit, we  determine  the  inflated error. This is a crucial step when dealing with combined datasets that have large systematic uncertainties. One  advantage of our method is that it  deals with  unknown systematic  errors of various experiments in a statistical manner.  As a byproduct, we provide some fits of the data that can be used elsewhere. A known disadvantage of using fits, is that  we may introduce unwanted structure. We are fully aware of this weakness and we do not draw strong conclusions based on the shapes of the functions used 
 in our fitting procedure. On the other hand, the obtained  spectral functions of the pion form factor suggest a complex non-perturbative structure rather than  clear  evidence of individual mesonic excitations.         

\section{ Fit for $\sigma_h$ and the inflated error}
\label{infl}

Mutual incompatibilities in data from various experiments are indicative of systematic errors. The methods used to determine experimental cross sections differ greatly. Uncertainties can originate from a new method that is not yet widely used. Conversely, examples of long-period systematics are also known. The determination of the neutron lifetime, the charged kaon mass, and the measurement of the Newtonian constant are known examples \cite{TRAMAX2025,ERHER2020} .    Systematic uncertainties, presented at measurement at continuous intervals (with the pion form factor being an example), complicate standard $chi^2$ analysis when applied to combined data.

  In the present paper the experimental error propagation is facilitated by the introduction of semi-stochastic inflated error (IE).
  When an inflated error ${\bar \sigma_I}$, is appropriately defined, it can serve as a partial accounting of systematic error, even when the error is not well-determined, in a statistical sense.  Given the absence of a consensus definition in the existing literature, we propose a novel approach to define IE through the fit  $\sigma_{fit}$  of data $\sigma_h$ :  the inflated error ${\bar \sigma_I}$ is defined such  that we get exactly (by requirement) $\chi_I^2=1$ in a theoretically allowed space of the cross sections $\sigma_{fit}(E) $ and inflated error functions ${\bar \sigma}_I(E)$  through minimization of the following quantity
\be \label{chicko}
\chi_c^2=\sum_j \frac{[\sigma_{h_c}(E_j)-\sigma_{fit, c}(E_j)]^2}{{\bar \sigma_{I,c}}^2(E_j)} \, .
\ee
where $c$ stands for a given hadron production channel and $E_j$ is the total energy of positron-electron pair.

In other words, we naively  assume that the systematic errors of sufficiently independent experiments can be considered as additional noise. 
Since the introduced IE replaces the covariance  error  matrices $\sigma^{stat}_{jk}$ of a single experiment, the following condition
 \be \label{ineq}
 \sigma^{stat}_{jk}\delta_{jk}<{\bar \sigma_I}(E_j)
\ee 
should be applied to account for the systematic error in the statistical sample of the combined data for given process.  
  Inequality  (\ref{ineq}) ensures that the inflated error can be taken equal to the  
 statistical error when a single experiment is considered. 
 Ideally, we should compare data with similar statistical errors to avoid proliferating experiments with low statistics. 
 Otherwise, we may undervalue  high-quality data, or, conversely, we may waste good data.
The latter can occur when excessive filtering according to the condition  (\ref{ineq}) is used.
  To avoid wasting data in a non-ideal case, the IE can always be chosen  larger, especially if its size is sufficient for the intended purpose. In this paper, we follow this approach and try to avoid excessive filtering.     
 
A simplified choice is used and we assume normal distribution for  inflated errors
\be \label{maj}
{\bar \sigma_{I,c}}(E)=N_c \sqrt{\sigma_{fit,c}(E)} \, ,
\ee
with channel (c) dependent, but energy-independent constant $N_c $.

\section{Dispersion relation for the pion electromagnetic charge form factor and the  $\pi\pi$ contribution to $\sigma_h$ }
\label{DRFOR}

Determining the cross section of the process  $e^+e^- \to \pi^+\pi^-$ cross section is of central importance.  It provides non-perturbative 
information about the distribution of quarks and gluons inside the lightest hadron. Through the dispersion relation for HVP, it also provides significant (75 $\%$) contribution to the muon anomalous magnetic moment.

The pion electromagnetic form factor $ F$ is  defined by the matrix element
\be
<\pi^+(p_+)|J^{\mu}|\pi^-(p_-)>=(p_++p_-)^{\mu}F(q^2) 
\ee
where the momentum $q=p_--p_+$. For the production process $q^2>4 m_{\pi}^2>0$, while for the spacelike 
electromagnetic form factor the kinematics $q^2<0$. The former is measured in the production processes for which we have collected data, the latter
is measured in electron-pion scattering experiments \cite{pispace} and pion production of nucleus  \cite{JLAB2001,JLAB2006,JLAB2007,JLABb2008}. In the timelike region, the modulus of the form factor appears in the   measured  hadronic cross section 
\be \label{pionform}
\sigma_{\pi\pi}(s)=\frac{\alpha^2(s)\pi\beta^3}{3s} |F(s)|^2 \, .
\ee    
where $s=q^2$.
Some experimental groups published the so called bare cross section $\sigma^B$ with HVP removed, i.e.
\be  \label{bar}
 \sigma^B(s)=|\alpha(0)/\alpha(s)|^2\sigma(s)
 \ee
where HVP was extracted on the basis of current  knowledge. 

In this work, however, we distinguish two form factors , the standard one $F$, and  the full one $F^F$, which is used  when the HVP and  other possible contributions are not removed from the cross section. In the single photon exchange approximation  we define  
\be
F_F(s)=F(s) |\alpha(s)/\alpha(0)|^2 
\ee
for which we can write the
\be \label{pion2}
\sigma_{\pi\pi}(s)=\frac{\alpha^2\pi\beta^3}{3s} |F_{F}|^2(s) \,. 
\ee

 It is not difficult to show that the full form factor $F_F$ satisfies the dispersion relation 
\be   \label{pionspec}
F_F(s)=\frac{1}{\pi}\int_{4 m_{\pi}^2}^{\infty} \frac{\rho_F(a)}{s-a+\ep}\, .
\ee
if the form factor  $F(s)$ obeys the same (with the replacement $ \rho_F(a) \rightarrow \rho(a)$).
In this work we  explicitly check that this is the case and the absorptive part of electromagnetic pion form factor,
 $Im F = -\rho$, is the sole quantity required to determine the function $F$ in the timelike and  the spacelike momenta as well. It can be extracted from lattice \cite{GADUED2024} or from  QCD equations of motion  \cite{VS2022}, however with limited precision only.
The fit for $\rho_F $ is therefore employed here. 

 For this purpose we chose to modify the well-known Gounaris-Sakurai fit, which is  particularly suited for this task \cite{pipiBABAR2012,SAHVP1, SAHVP2}. 
However, since the original model \cite{GOSA1968} violates unitarity, we improved upon it, and used its  unitarized and analyticized version. The resulting amplitude  satisfies the 
dispersion relation with standard cut along the real axis of the variable $q^2$, where the  discontinuity   is identified with the imaginary part of the original Gounaris-Sakurai model.   
 
 Explicitly written, the spectral function  has the following form:
\be \label{mainfit}
\rho_F(a)= \Im \frac{1}{\cal N} \left[{\cal W}^{GS}_{\rho}(a,m_{\rho},\Gamma_{\rho})(s)\frac{1+c_{\om}{\cal W}_{\om}(a,m_{\om})}{1+c_{\om}}
+c_{\phi}{\cal W}^{GS}_{\phi}(a)
+\sum_{i}c_i {\cal W}_{GS}^i(a,m_i,\Gamma_i)\right]\, \, .
\ee
where $c$ are real coefficients and  two parametric function ${\cal W}^{GS}_{i}$  and ${\cal W}$ are listed in Appendix \ref{appendix:fit}.

 The real part  of the function $F_F$  is then calculated through (\ref{pionspec}), substituted into the cross section \ref{pion2} in case of full form factor consideration (as was done, for example for the  set with CMD-3 data included in).  Equivalently, using  $\sigma_B$ instead (since provided by other experimental groups) we fit the form factor $F$ instead by the  formula (\ref{mainfit}). The latter approach was followed when the combined world data set was used.   In a strict sense, full cross-section data are necessary for a new fresh HVP extraction. Using bare data always relies on the HVP extracted in the past.  In this paper, we ignore the secondary error of not adding HVP known at the time of the given measurement, but it is done iteratively by solving the Eq. (\ref{muf}) as an integral equation.

This  fit provides the inflated error determined by the constant  $N^2=0.49~nb$ in Eq. (\ref{maj})  which   is  slightly better than  obtained previously  \cite{SAHVP1, SAHVP2}.
 Here we did not refer \cite{sauliarxiv}, e.g. due to the later identification of the error made 
 in the code for the pion form factor evaluation.  The error is corrected in this paper (see Appendix \ref{appendix:trick}).

The fitted parameters are listed in Table \ref{tabrho}. Reducing the IE further would require filtering  the data. Since the achieved error is sufficient for determining error propagation, we will stop here. The  results obtained suggest that the inflated error method represents a meaningful method in many respects. Although extracting errors is a time-consuming task, the benefit of easy later use is significant.

 Conversely, despite achieving a precise and smooth analytical fits, no strong conclusions should be drawn from the resulting fits of both data sets. Note that half of the potential rho-meson excitations are ghosts, according to their negative couplings, $c_i$, and no such structure could straightforwardly be identified as a radial rho-meson excitation.  These ghosts overlap normal sign broad resonances and  are  needed to  shape the intrinsic non-perturbative effects reflected in the amplitude. While negative couplings are  accepted and used by experimental groups, there is no meaningful low energy theory  to interpret  them.

Perhaps more importantly, there is no evidence of  of analyticity distortion due to systematic experimental error.  The CMD-3 data-governed fit with standard statistical deviations has achieved a chi-squared value of 1, indicating  that the dispersion relation (\ref{pionspec}) is excellently fulfilled in this case.  
 We are fully aware of the  existing unitarity / analyticity bounds \cite{simula}, which, in principle can provide a fit-independent test of our analytical assumptions. However, these methods are very difficult to apply to the timelike combined data considered here. To perform  these analyses, we wrote two independent codes, one uses Python libraries and  the other  is based on repeated iterations. 
 For the $e^+e^-\to \pi\pi$ experimental cross section we  used the data collected by the KLOE \cite{pipiKLOE2005} by  CMD/SND operating group  \cite{pipiCMD2005}, SND \cite{pipiSND2006}  detectors as well as data extracted  by  BaBar \cite{pipiBABAR2012}, and BESS-III \cite{pipiBESSIII}.

\begin{figure}[htb]
\includegraphics[width=9.cm]{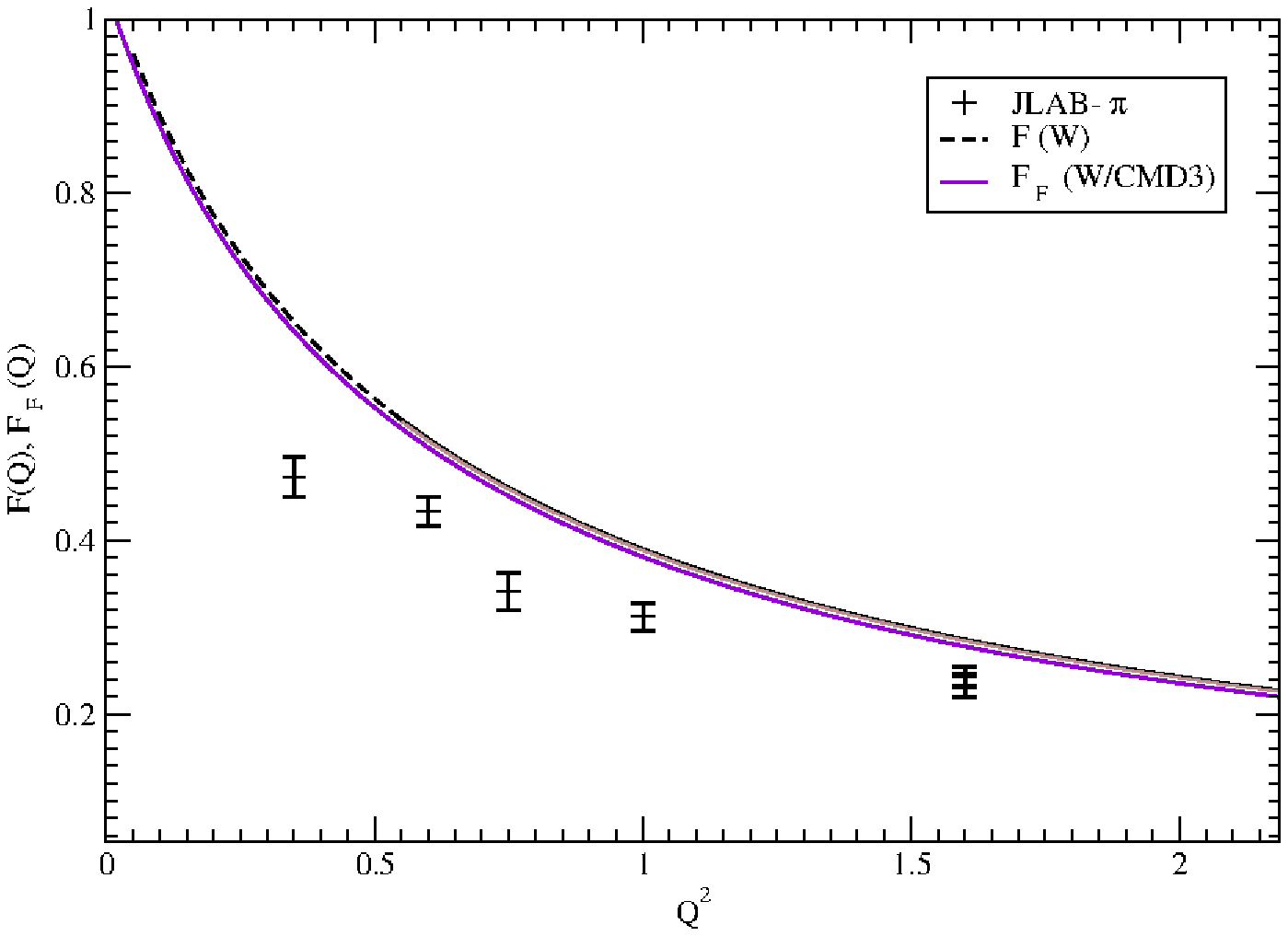} \includegraphics[width=8.cm]{JLAB.eps}

\caption{\label{space} Pion electromagnetic form factors continued from the timelike to the spacelike region. Left panel: 
Pure QCD pion  form factor $F$ (dashed) as obtained from (W) world combined data without CMD-3 data and the full form factors $F_F$ (solid) as extracted from the data that includes CMD-3.
The error band is shown for the later case.  The upper bound is shown only from the $Q^2=0.5 {\mbox GeV}^2$  for purpose of visibility. Right panel:  Comparison of  form factors with  PT QCD asymptotics. W (CMD3) indicate sets of data used to get the form factor.}
\end{figure}

For the data governed by the  CMD-3 experiment the $n_{dof}=209+45$, where the latter number in the sum is for remaining data outside the CMD-3 range.   To  show the error band, like  in Fig. \ref{space}, the  propagation of error was estimated within use of inflated error, where we took $N^2=0.49$ nb.  We calculated  the individual inflated errors for each channel in the HVP calculation separately.  However, this level of precision turns out to be unnecessary for correctly determining  error propagation and for the purposes of this paper we use a common inflated error ($N^2=1$ nb) when calculating the error bands in the  figures presented.
\\
\begin{figure}[htb]
\centerline{\includegraphics[width=7.5cm]{nufik4.eps} {\mbox , } \includegraphics[width=7.5cm]{nufik.eps}}
\caption{\label{nufik} Left:  Spectral functions $\rho_F$ ($\rho=\rho_{\pi}$)  of the pion form factor $F_F$ ($F$) as extracted from data with CMD-3 data and without $CMD-3$ data.  Right: Examples of selected  cross sections as described in the text. The full cross section $\sigma_F$ fit (solid) governed  by  CMD-3 data is shown. The bare data-set with CMD-3 data excluded and the associated fit of bare cross section (dashed) are  shown as another example. Deviations are not shown for purpose of visibility. }
\label{figure5}
\end{figure}

The QCD fitted spectral function $\rho$  is shown in Fig. \ref{nufik}. The set that do not contain CMD-3 data was used in this case. Fits needed in our study are shown on the right panel. The difference between cross sections are due to vacuum polarization presence/absence (absent for $\sigma_B$) and due to the systematics as well. For further  comparison of CMD-3 and other datasets we refer to the original experimental paper \cite{CMD3}.

\begin{center} 
\begin{table}
\begin{tabular}{ |c|c|c|c|c|c|c|c| }
\hline
fit                           &   \multicolumn{3}{c|}{with CMD-3 data} && \multicolumn{3}{c|}{ W (world without CMD-3)} \\
\hline
name                    &  m           &$\Gamma $&  c                    &&   m         & $\Gamma$ &  c         \\
\hline 
 $ 0$               &  0.7219   &  0.206       & -0.0518                  && 0.6895    &   0.113        &  -0.036     \\
$ \rho $               & 0.77453   & 0.15173    &  1                          && 0.7754   & 0.1565         &  1               \\
$\omega$           & 0.7821 &   0.0086099   &  0.001754          && 0.7817783   & 0.0087933 &  0.00169      \\
$\phi $                &1.02181    & 0.00459          &0.0007      && 1.02144    &   0.0042      &  -0.0008       \\
 $ 1 $              & 1.3832       &   0.291           & -0.0609          &&  1.403   &  0.206             &   -0.0597        \\
 $ 2 $            &  1.694     &    0.123          & 0.0161                &&  1.684    &  0.104            &  0.0196            \\
 $ 3 $          &  2.45         &     0.66       & -0.0153                  &&  2.28     & 0.49             & -0.017           \\
 $ 4 $          &  -         &     -        & -                                    &&   2.25          &  0.1            &  0.0047          \\
 \hline
\end{tabular}
\caption{\label{tabrho} List of parameters  for fits of pion electromagnetic form factor $F_F$ based on analyticized Gounaris-Sakurai model with  CMD-3 data (left part), and the same for $F$
for world data without CMD-3 data (right part). 
Normalization prefactors:  ${\cal{N}}_{CMD3}=0.8539$; ${\cal{N}}_{W}=0.8615$. }
\end{table}
\end{center}

\subsection{Results for the pion electromagnetic form factors}

Using the obtained  spectral functions the continuation to the spacelike region of momenta is straightforward. Continuations based on two extracted spectral functions are equally compatible with the experimental JLAB-$\pi$ data. This can be seen in Fig. \ref{space}, 
where comparison is made. In fact the CMD-3  data-driven result for the electromagnetic form factor turns to be even slightly closer to the JLAB measurement. To determine the form factor,  the statistical errors provided by CMD-3  were used, while the inflated error was used to calculate the function  $F$ with the spectral function $\rho$ based on the combined world data.

Another interesting point is the question of agreement with perturbation theory (PT) QCD, where a known formula $F\rightarrow \beta Q^{-2}\ln^{-1}(Q^2/\Lambda^2) $ has been obtained \cite{CZ1977,FJ1979,LB1980,ER1980}. Assuming the validity of the dispersion relation \ref{pionspec}, we find no evidence of PT asymptotics from in any data set. 
The  PT  form factor is compared with analytically continued 
form factors in the right panel of Fig. \ref{space}. The asymptotic  slopes of the continued form factors differ   significantly  from the known PT form, showing   no evidence of  the inverse log behavior. 

We did not include systematic errors resulting from theoretical assumptions, nor did we speculate on what future data could be.  However, as was shown in a recent paper on sum rules, using the PT QCD dispersion relation would not improve the situation. According to the paper \cite{ARPUWE2025}, we also note that  recovering PT QCD asymptotics  within the framework of dispersion relations is  unlikely, irrespective of future experimental  access to higher $q^2$.

 Even more interestingly, the unpublished JLAB data show possible evidence of an ongoing $1/Q^2$ asymptotic.  It was actually observed that   $Q^2 F(Q^2)$ takes a constant value for $Q$ in several ${\mbox GeV }$. Theory for  exclusive QCD processes should be revisited (e.g., the failure of factorization could be explained). Alternative explanations based on the fine-tuned  interplay between hadronic resonances and PT QCD  are sought \cite{ARPURBE2026}.

 \section{Results for $\alpha_{QED}$ at KLOE-2}
\label{alfy}

The QED running coupling is an observable that inversely depends on  HVP. Two analytic fits for the cross section $\sigma_{\pi\pi}$ as described in the previous sections, were employed to  compare with the KLOE-2 data for the running coupling.  
 To determine HVP, in  addition to the cross section $\sigma_{\pi\pi}$, we use the known  fits \cite{sauliarxiv,SAHVP1,SAHVP2}  for the remaining exclusive channels, including $K^{+}K^{-}$, $K_LK_S$ and $\pi\pi\pi$, as well as subdominant $\eta\gamma$ and $\pi\gamma$ production cross sections in order to determine HVP through the formula (\ref{muf}). All  fits to aforementioned experimental data  were made and   summarized in \cite{sauliarxiv} and they are used without modification. 
Final states with four pions were included, while neglecting $KK\pi$ and the other states with a multiplicity greater than three.  Furthermore, the well-established vector charmonia and bottomonia were included in  $\sigma_h$ via their BW forms with PDG-averaged values, following the standard routine.

 The integral equation (\ref{muf}) was solved iteratively using  a  large numerical grid to ensure that the numerical error was much smaller than the associated statistical and systematic deviations. The codes and resulting data are available to the public at pages of NPI institute. As in the case of the pion electromagnetic form factor, the  numeric were enforced  by using the so-called Hollinde trick when  performing  principal value integration.

We compare with the  QED running coupling, which has  been measured below 1 {\mbox GeV}  by the KLOE-2 collaboration \cite{KLOEdva}. Two considered calculated results for the QED running coupling are compared in  Fig.~\ref{celkove}  for the KLOE-2 accessible region.

Two determined HVP functions provide two distinct results for the running coupling. 
The incompatibility between the data sets with and without the CMD-3 data is too small compared to the experimental errors of KLOE-2 measurements. The same argument applies in the case of different results obtained by our method and by others \cite{polar2,polar3}.

{\mbox{}}
\\
\begin{figure}[ht]
\centerline{{\includegraphics[width=7.60cm]{global1.eps}}}
\caption{Square of the fine structure constant as affected by the presence or absence of CMD-3 data for the
$\sigma_{\pi\pi}$ cross section. The line labeled by F.J.  stands for HVP extractions made by \cite{polar2}.}
\label{celkove}
\end{figure}
\\
{\mbox{}}

\section{Conclusion}

We examined the implications of incorporating CMD-3 data and assessed the pion form factor within the time-like and space-like domains of momenta. Additionally, we evaluated the running fine structure constant at low momentum and compared it with experimental results. 
 Despite the fact that the experimental points for $\sigma_{\pi\pi}$ provided by the CMD-3 group  represent  an outlier case among the  other experiments, we have shown that  the incompatibility imprints in related observables do not cause further tension between theory and experiments.

  The first observable discussed is the elastic scattering of a pion on an electron (or proton).  
 It is a matter of fact that the spacelike continuation of the CMD-3-driven electromagnetic form factor is slightly closer to the experimental values.  Additionally, we estimate the propagation of errors to a large spacelike $Q$ and observe little chance of 
 achieving the naive perturbative QCD limit. Assuming the form factor  satisfies the usual dispersion relation, the function $F$ in the deep spacelike region of momenta is unlikely to ever meet the PT QCD prediction. 
   
Furthermore, the obtained spectral functions, together with the other considered spectral fits for hadronic production cross sections, was used to calculate the QED running charge in the time-like region of the momenta and  with the 2016 KLOE-2 measurement for the electromagnetic running charge.  In this case it would require at least  ten times better precision than the KLOE-2 experiment to potentially detect  incompatibility of the data. Similar comparisons as presented in our study  can be  a challenging task for future high-precision measurements.


\appendix

\section{Details of the analytical fit}
\label{appendix:fit}

In this appendix we review the list of functions used in our constructions.

Particularly important is the form of the Gounaris-Sakurai dressed vector meson propagator, which is
\be  \label{GSprop}
{\cal W}^{GS}=\frac{m^2+d(m)\Gamma/m}{M^2(s)-s-i m\Gamma(s,m,\Gamma)} \, \, ,
\ee
\be
M^2(s)=m^2\left[1+\frac{\Gamma k^2(s)}{k^3(m)}(h(s)-h(m^2))+\frac{\Gamma h^{'}(m^2)}{k(m)}(m^2-s)\right] \, \, ,
\ee
\bea
\Gamma(s,m,\Gamma)&=&\Gamma\frac{m}{\sqrt{s}}\left[\frac{L_2(s,m_{\pi})}{L_2(m^2,m_{\pi})}\right]^3 \, \, ,
\nn \\
L_2(s,m_{\pi})&=&\sqrt{s-4m^2_{\pi}} \, \, , \label{eldva}
\eea
where we have defined following auxiliary functions:
\be
h(s)=\frac{\beta(s)}{2}\ln\left(\frac{\sqrt{s}+2k(s)}{2m_{\pi}}\right) \, \, ,
\ee
\be
 h^{'}(m^2)=\frac{2 m_{\pi}^2 h(m)}{m^4\beta(m)}
+\frac{2 m_{\pi}^2}{\pi m^4 \beta(m)}
+\frac{\beta(m)}{2\pi m^2} \, \, ,
\ee
\be
d(m)=\frac{4m_{\pi}^2}{m^2\beta^3(m)}(3h(m)-2/\pi)+\frac{1}{\pi\beta(m)} \, ,
\ee
with the usual shorthand notation:
\be
\beta(s)=\frac{L_2(s,m_{\pi})}{\sqrt{s}} 
\, \, ,\, \, 
k(s)=\frac{L_2(s,m_{\pi})}{2} \,\, ,
\ee
 used for the velocity and the  two pion Lorentz invariant phase space factor.

 The Breight-Wigner function for the narrow $\omega$ meson was taken in the form:
\be
{\cal W}_{\omega}=\frac{m^2_{\omega}}{m^2_{\omega}-s-im_{\omega}\Gamma_{\omega}} \, .
\ee     
 
 \section{Holinde trick}
\label{appendix:trick}

  To improve the  principal value integration numerically we subtract (add) the exact zero from the original integral 
 \bea
  \int_T^{\infty} d x \frac{f(x)}{y-x}= \int_T^{\infty} d x \frac{f(x)(x+y)-f(y)2y}{y^2-x^2}
  \nn \\
  +f(y)\ln\frac{1-T/y}{1+T/y}
  \eea
  , which is the actual form used for discretization to the sum.
  
  It is the term in the second line that has been neglected in the evaluation of the pion form factor
  in the unpublished paper \cite{sauliarxiv}.


%
\end{document}